# Bayesian Analysis of Target Detection with Enhanced Receiver Operating Characteristic

Philip Cassady, PhD

Cassady Engineering Company

pcassady@alum.mit.edu

**Abstract:**

In the present paper we develop a Bayesian analysis of radar target detection that uses the parameters of conventional radar analysis to provide a valid prediction of target presence or absence when received signals cross or fail to cross chosen threshold values. A Positive Predictive Value parameter is added to the normal Receiver Operating Characteristic to provide information that allows the radar operator to make an informed decision in the choice of threshold.

**Introduction:**

Bayesian statistical analysis provides a method to incorporate new data into existing prior probability calculations to provide an improved posterior probability value. The detection of radar pulses using an Inverse Probability Receiver is described in radar systems textbooks (1). The method has been applied by Woodward and Davies (2, 3, 4) to the reception of radar signals in the presence of white noise. Woodward and Davies use information theory to elicit the details of a transmitted waveform from the combination of transmitted signal and white noise in the received waveform. The use of detection theory including Bayesian analysis in the field of psychophysics is extensive as described by Swets (5). In the present paper we develop a Bayesian analysis of radar target detection that uses the parameters of conventional radar analysis to provide a valid prediction of target presence or absence when received signals cross or fail to cross chosen threshold values. A Positive Predictive Value parameter is added to the normal Receiver Operating Characteristic to provide information that allows the radar operator to make an informed decision in the choice of threshold.

**Analysis:**

Conventional radar signal analysis is presented in terms of the probability of detection, Pd and the probability of false alarm, Pfa. The probability of detection is the probability that the target will be detected given the fact that the target is truly present. The probability of false alarm is the probability that the target will be "detected" by the measurement when in fact the target is not present. In the notation of conditional probability:

> Pd = Probability (Measurement will be positive|Target is truly present)

> Pfa = Probability (Measurement will be positive|Target is not present)

It would be much more useful to be able to turn the definition of Pd around to obtain the so called Posterior probability, the probability that a target is truly present when the measurement is positive:

> Posterior = Probability (Target is truly present|The measurement is positive)

Bayesian analysis turns the conditional Pd probability around and provides a value for the probability that the target is in fact present when the measurement provides a positive detection signal. This information is much more valuable because it provides the probability that a target is present if a positive detection signal is received. To perform this Bayesian analysis, however, it is necessary to assume a so called Prior probability that the target is present before the measurement is made. The necessity of this Prior probability assumption has been a source of controversy concerning the Bayesian analysis. This Prior is merely a guess concerning the presence of the target that is made prior to the actual measurement. If one does not know a-priori whether or not a target is present, it is certainly acceptable to assume an uninformed 50 percent Prior probability, equal odds that the target is present or not.

The method is best understood by generating a table that includes the effects of the Prior probability together with the Pd and Pfa of the new measurement data. Here we are assuming that we have a prior estimate of the target presence, here called the Prior, and we have taken a new measurement using a method that has its own intrinsic Pd and Pfa. We want to derive a new Posterior probability that combines both the original prior estimate with the new measurement data.

|  | Target is truly present | Target is not present | Totals |
| --- | --- | --- | --- |
| Detection is positive | Pd * Prior | Pfa*(1-Prior) | Pd*Prior+Pfa*(1-Prior) |
| Detection is negative | (1-Pd)*Prior | (1-Pfa)*(1-Prior) | (1-Pd)*Prior + (1-Pfa)*(1-Prior) |
| Totals | Prior | 1-Prior | 1 |

**Table 1.**

This matrix can be displayed in terms of a logic matrix:

|  | Target is truly present | Target is not present |
| --- | --- | --- |
| Detection is positive | True positives | False positives, False Alarms |
| Detection is negative | Missed detections | True Negatives |

**Table 2.**

A positive detection is defined as a return that is larger than the chosen threshold, and a negative detection is defined as a return that is less than the threshold

The Posterior probability that the target is present after a positive, above threshold, detection signal return will be equal to the number of positive detections when the target is truly present divided by the sum of the number of positive detections when the target is truly present and the number of positive detections when the target is not present (False Positives):

    Posterior = Pd*Prior / [Pd*Prior + Pfa*(1-Prior)]     (1)

The logic of this Posterior is clearly seen using the information from the logic matrix, Table 2.

Posterior = Probability that a Target is truly present after receiving a positive return signal =

    = Number of True Positives/(Number of True Positives + Number of False Alarms) =

    = Number of True Positives/Number of All Possible Positives

The relationship given in Equation 1 is quite simply stated in terms of the odds and the Likelihood:

    Posterior Odds = Likelihood * Prior Odds     (2)

where

    Posterior Odds = Posterior/(1 - Posterior)     Prior Odds = Prior/(1 - Prior)     (3)

    Likelihood = Pd/Pfa     (4)

The Posterior is adversely affected by both a low value of the Prior and a large value of Pfa for the measurement as shown in Figure 1.

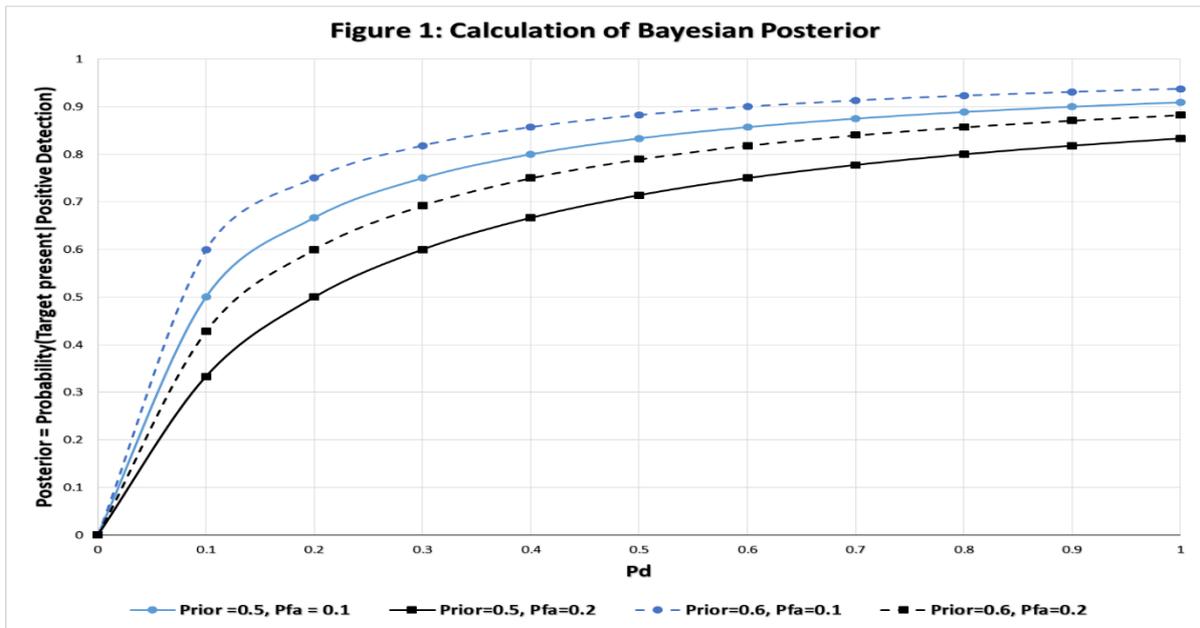

As a starting point, one can assume a uniform, non-informed Prior probability that the target is present, a Prior = 0.5 (equal odds that the target is present or absent). The results are shown by the solid curves in the figure. For a given measurement the Posterior probability increases with increasing values of the measurement Pd, and decreases with increasing values of the measurement Pfa, as one would expect. If one is able to increase the initial estimate of Prior probability to a value of Prior = 0.6, the Posterior probabilities are improved considerably, as one would expect.

After the first measurement is performed, the accuracy of the Posterior can be improved by incorporating another, second measurement in the Bayesian analysis. If the first measurement return was positive, then the Posterior is, in fact, the probability that the target is present and can be used as a Prior estimate for a Bayesian analysis using a second measurement. The effects of several sequential measurements can be easily calculated using this analysis.

Examples of sequential measurements are shown Table 3. In all cases the Prior for the first measurement is taken as 0.5 (equal odds that the target exists or not). The values of Pd and Pfa for the first measurement are as shown. The calculated Posterior for the first positive measurement is shown as Posterior 1. This Posterior 1 is then used as a Prior 2 for the second measurement, which is assumed to have the same values of Pd and Pfa as the first measurement (same detector threshold settings used for both measurements). The calculated value for the Posterior for this second positive measurement is considerably improved, especially for low values of Pd and high values of Pfa.

| Prior 1 | 0.5 | 0.5 | 0.5 |
| --- | --- | --- | --- |
| Pd | 0.9 | 0.7 | 0.7 |
| Pfa | 0.01 | 0.01 | 0.1 |
| Posterior 1 = Prior 2 | 0.989 | 0.986 | 0.875 |
| Posterior 2 = Prior 3 | 0.994 | 0.993 | 0.925 |
| Posterior 3 | 0.944 | 0.977 | 0.804 |

**Table 3**

If the measurement return was negative, below threshold, then the probability that the target is present given that the measurement was negative is, using the Table 1:

Probability (Target is present | Measurement was negative) =

$$(1-Pd)*Prior / [(1-Pd)*Prior + (1-Pfa)*(1-Prior)] \quad (5)$$

This probability is used as the Prior for the Bayesian analysis used for the negative measurement.

The logic of this calculation is clearly seen using the information from the logic matrix, Table 2:

Probability that a target is truly present after receiving a negative return signal =

= Number of Missed Targets/(Number of Missed Targets + Number of True Negative Targets)

= Number of Missed Targets truly present/Total Possible Number of Negative Returns

The relationship given in Equation 5 can be quite simply stated in terms of the Odds given in Equation 3 and the Complementary Likelihood:

Posterior Odds = Complementary Likelihood * Prior Odds  (6)

Where

Complementary Likelihood = $(1 – Pd)/(1 – Pfa)$  (7)

The effect of a negative measurement is shown in Table 3, assuming that a third measurement returned a negative result. The scenario is: Assuming an initial uniform Prior = 0.5, three measurements were taken. The first two measurements returned positive detections, and the third measurement returned a negative detection.

Table 3 shows that the first positive detection raises the probability that a target is present from the initially assumed value of 0.5 to a value near 0.9 depending on the Pd and Pfa performance characteristics of the detector. The second positive detection further raises the probability that a target is present. The third measurement, which returned a negative detection, lowers the probability that a target is present in all cases.

**Multiple Sequential Measurements:**

It is possible to derive an expression for the cumulative probability that a target is present In the normal situation when a detector is used to take sequential looks for a target using the same values of Pd and Pfa for all of the sequential looks (the target statistics and detection threshold are held constant for all of the subsequent looks). In this situation the posterior probability of each look is used as the prior probability for the next subsequent look. Using Equation 1 for any look that returns a positive return and using Equation 5 for any look that returns a negative return the expression for the cumulative probability that a target is present after N looks is:

Probability that target is present after N looks =

$$\text{Posterior}_N = \frac{1}{1+\left(\frac{Pfa}{Pd}\right)^{N_P}\left(\frac{1-Pfa}{1-Pd}\right)^{N_N}\left(\frac{1}{Prior}-1\right)} \quad (8)$$

Where Prior is the original Prior, usually taken as 0.50, $N_P$ is the number of positive threshold crossings, and $N_N$ is the number of negative (failed) threshold crossings.

Again the relationship given in Equation 8 is quite simply stated in terms of the Odds given in Equation 3 and the Likelihoods given in Equations (4) and (7):

Posterior Odds$_N$ = (Likelihood)$^{Np}$ * (Complementary Likelihood)$^{NN}$ * Prior Odds  (9)

Individual measurements with different values of Pd and Pfa can be accommodated using the corresponding individual values for Likelihood and Complementary Likelihood and appropriate values for Np and NN.

**Receiver Operating Characteristic Enhanced with Positive Predicted Value:**

The radar receiver operating characteristic, ROC, is a plot of the probability of detection, Pd on the vertical axis against the probability of false alarm, Pfa, on the horizontal axis. Curves of constant SNR are plotted with the threshold as a parameter. The ROC curve is generated as the detection threshold is varied from a low value (high Pd, high Pfa) at the upper right corner of the ROC to a high value (low Pd, low Pfa) at the lower left corner of the ROC. The radar operator must choose the threshold value to provide the best trade-off between the conflicting requirements for adequately high probability of detection, Pd, and acceptably low probability of false alarm, Pfa.

The Bayesian Posterior, called the Positive Predictive Value, PPV, (5) gives the probability that a detection will truly indicate that a target is present. It depends on both the Pd and Pfa and also the assumed Prior probability that a target is present.

The PPV is given by the same equation as the Posterior (Equation 1):

$$\text{PPV} = \text{Pd}*\text{Prior}/[\text{Pd}*\text{Prior} + \text{Pfa}*(1-\text{Prior})] \qquad (10)$$

The various terms can be understood with reference to two pdf's that are common to radar target detection and presented in Figure 2, which is drawn for Rayleigh Noise pdf and Rician Signal+Noise pdf:

$$\text{Rayleigh pdf} = \frac{x}{\sigma^2}\exp\left(-\frac{x^2}{2\sigma^2}\right) \qquad \text{Rician pdf} = \frac{x}{\sigma^2}\exp\left(-\frac{x^2+s^2}{2\sigma^2}\right)I0\left(\frac{xs}{\sigma^2}\right) \qquad (11)$$

And s/σ is the SNR

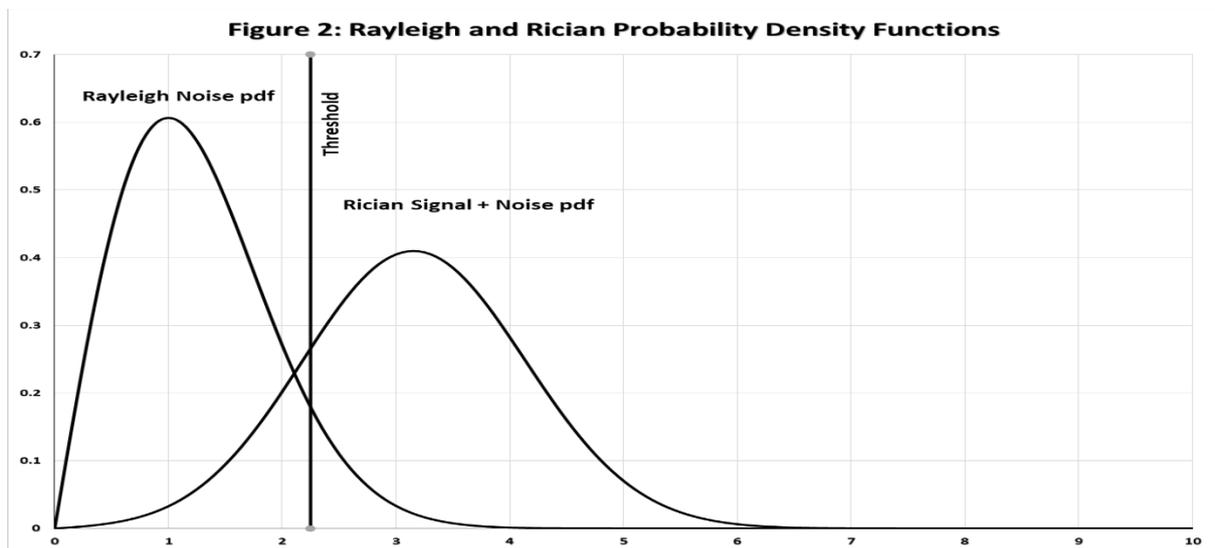

Figure 2: Rayleigh and Rician Probability Density Functions

The Pd is represented by the area under the Rician pdf to the right of the Threshold. The Pfa is represented by the area under the Rayleigh pdf to the right of the Threshold.

The analyses presented here can be applied to other fields including infra-red search and track and medical decision applications with the use of appropriate pdf's, which are usually Gaussian, for these fields

Using these definitions, the Receiver Operating Characteristic including the PPV shown in Figure 3 has been calculated for a SNR value of 2 and a non-informed uniform Prior of 50 percent:

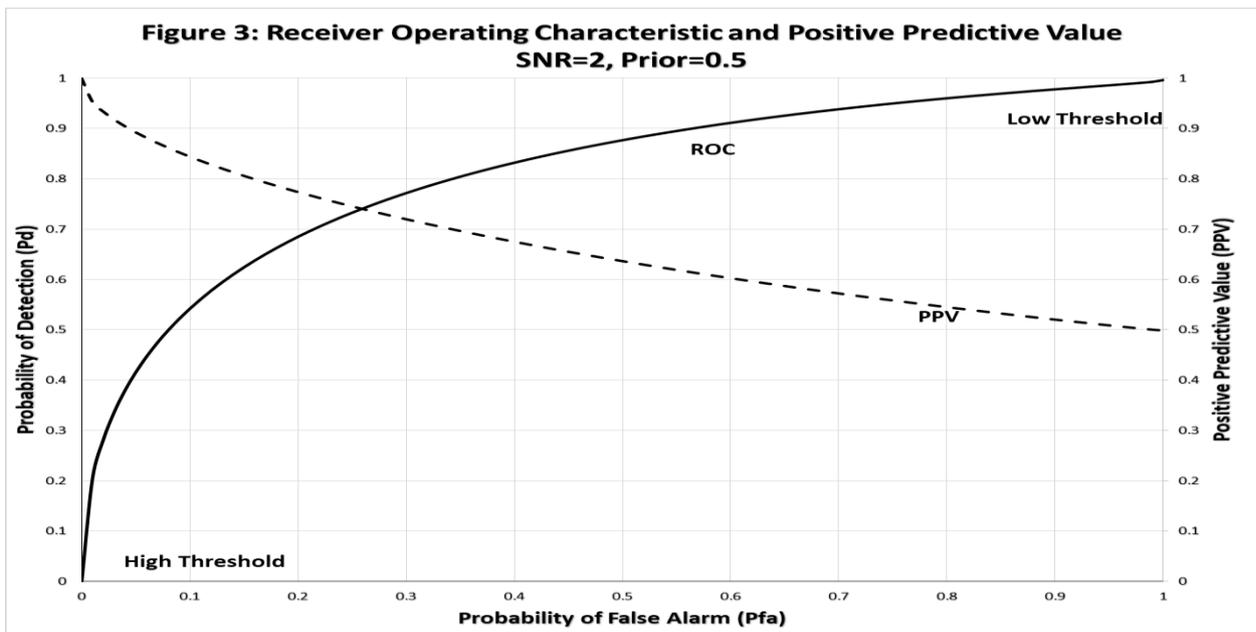

The inclusion of the PPV curve gives the radar operator an actual probability of true target presence with which to choose the threshold value.

For example, if the operator sets his threshold to provide a Pd of 0.8 he will also have a Pfa of 0.35. The accompanying PPV of 0.70 indicates that only 70 percent of his detections will be true detections. If he increases his threshold to lower the value of Pd to 0.5, he will have a better Pfa of 0.08, and the PPV will increase to 0.86 indicating that 86 percent of his detections will be true detections. By raising his threshold he will receive fewer detections, but more of them will be genuine.

The benefit of using the PPV enhanced ROC can be seen when the SNR increases from 2 to 3 as shown in Figure 4.

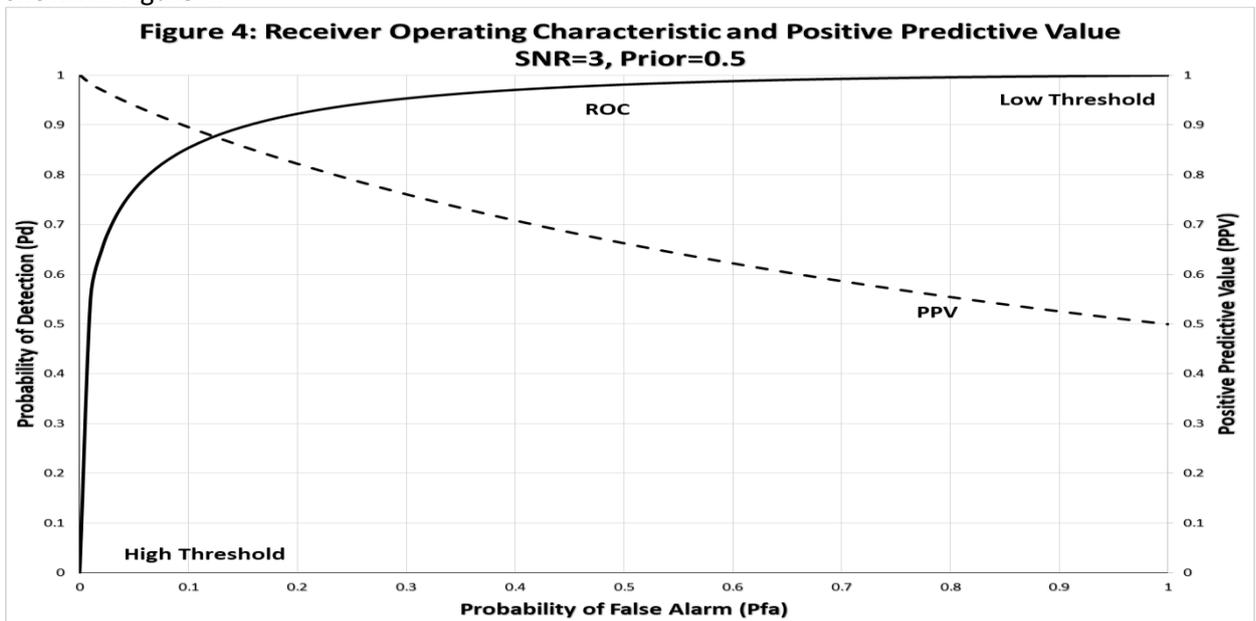

Comparison of Figure 4 with Figure 3 shows that if the operator wants to maintain the signal integrity (PPV) of 80% when the SNR increases from 2 to 3 he can decrease the threshold to the point that the Pfa increases from 15% when SNR=2, to 23% when SNR=3, this decrease in threshold will also cause the Pd will increase from 63% when SNR=2 to 94% when SNR=3. The large increase in Pd more than compensates for the small increase in Pfa when the threshold is decreased. Use of the Enhanced ROC with PPV allows the operator to make informed trade off between Pd and Pfa on the ROC curve.

**Conclusions:**

We have described a Bayesian analysis for signal detection in this paper as applied to the process of radar target detection. The basis of the analysis was described using a logic matrix. The resulting Posterior probability for target detection was calculated and shown to provide valuable information to the radar operator. A procedure was developed for a string of subsequent measurements to provide improved detection statistics. An enhanced ROC was developed that allows the radar operator to interpret the trade off between Pd and Pfa and to make an informed adjustment of the detection threshold to obtain optimum target detection. This analysis can easily be applied to other fields that require signal detection and decision with the use of appropriate pdf's for the application.

**Acknowledgement:**